\documentclass[10pt, conference]{IEEEtran}
\IEEEoverridecommandlockouts

\usepackage{cite}
\usepackage{amsmath,amssymb,amsfonts}
\usepackage{graphicx}
\usepackage{textcomp}
\usepackage{xcolor}
\def\BibTeX{{\rm B\kern-.05em{\sc i\kern-.025em b}\kern-.08em
    T\kern-.1667em\lower.7ex\hbox{E}\kern-.125emX}}
\usepackage{comment}
\usepackage{tagging}

\usepackage{graphicx}
\usepackage{wrapfig}
\usepackage{authblk}
\usepackage{textcomp}
\usepackage{xcolor}
\usepackage{wrapfig}
\usepackage{setspace}
\usepackage{color}
\usepackage[normalem]{ulem}
\usepackage{multirow}
\usepackage{hhline}
\usepackage{tikz}
\usepackage{url}
\usepackage{enumitem}
\usepackage{subcaption}
\usepackage{tikz}
\usepackage{kotex}
\usepackage{tabularx}

\usepackage{graphicx}
\usepackage{textcomp}
\usepackage{xcolor}
\usepackage{tcolorbox}
\usepackage[normalem]{ulem}
\usepackage{setspace}
\usepackage{colortbl}
\usepackage{xfp}

\usepackage[utf8]{inputenc}
\usepackage{kotex}
\usepackage{color}
\usepackage{algorithm}
\usepackage{algpseudocode}
\usepackage[normalem]{ulem}
\usepackage{makecell}
\usepackage{dblfloatfix} 
\usepackage{lipsum}
\usepackage{booktabs}
\usepackage{multirow}
\usepackage{siunitx}
\usepackage{verbatim}
\usepackage[colorlinks=true,linkcolor=violet,citecolor=violet,urlcolor=black]{hyperref}
\usepackage{circledsteps}
\definecolor{darkred}{RGB}{204,0,0}

\newcommand{\CircledTextBlack}[2][]{%
    \CircledParamOpts{inner color=white, outer color=none, fill color=black, inner xsep=2pt, inner ysep=2pt, #1}{1}{#2}%
}

\newcommand{\CircledTextRed}[2][]{%
    \CircledParamOpts{inner color=white, outer color=darkred, fill color=darkred, inner xsep=2.5pt, inner ysep=2.5pt, #1}{1}{#2}%
}

\usepackage{ulem}

\newcommand{\us}{$\mu s$}

\newcommand{\squishlist}{
\begin{list}{$\bullet$}
	{ \setlength{\itemsep}{0pt}      \setlength{\parsep}{-0pt}
		\setlength{\topsep}{4pt}       \setlength{\partopsep}{0pt}
		\setlength{\listparindent}{-2pt}
		\setlength{\itemindent}{-5pt}
		\setlength{\leftmargin}{1em} \setlength{\labelwidth}{0em}
		\setlength{\labelsep}{0.5em} } }

\newcommand{\squishend}{
\end{list}}

\newcommand{\tbdfixed}{\textsc{{CXDVirt}}}

\begin{document}

\title{\tbdfixed{}: Low Latency Kernel Module Based CXL-SSD Emulation}

\author{
{Hyunsun Chung}$^{1}$, {Seongho Bong}$^{1}$, {Hong-Yeon Kim}$^{2}$,
{Youngjae Kim}$^{1,\dagger{}}$
\thanks{$^{\dagger}$Y. Kim is the corresponding author.}
\\
{$^1$Sogang University, Republic of Korea, 
$^2$ETRI, Republic of Korea}
}

\maketitle

\begin{abstract}
CXL-SSDs promise memory-semantic access to NAND-scale capacity through a
small in-device DRAM cache, but scarce hardware makes emulation central to
CXL-SSD research. Cylon, a state-of-the-art emulator, runs workloads in a VM,
intercepting DRAM misses to inject NAND latency. We show this VM machinery
adds over 3\us{} per miss, exceeding high-performance NAND read latency, while
repeated VM exits inflate tail miss latencies several-fold. We present
\tbdfixed{}, a kernel-module-based CXL-SSD emulator that removes the VM
entirely: the device is exposed through host devdax, hits run natively via
the MMU, and a custom page-fault path models NAND latency on misses.
\tbdfixed{} preserves CXL-SSDs' bimodal latency profile while reducing
per-miss latency 1.4--1.6$\times$ (4.1$\times$ at P99.9); with eight threads
its miss latency only doubles, against 8--9$\times$ for Cylon behind QEMU's
global lock. \tbdfixed{} stays within 5\% of remote DRAM without NAND traffic, where Cylon is 3.1--3.6$\times$ slower, and enables configurable eviction and prefetch policy studies.

\end{abstract}


\section{Introduction}

The growth of memory-intensive workloads in AI and HPC alike has driven increased interest in memory expansion related research in academia and industry~\cite{gholami2024memorywall, kwon2018memorywall, yang2026beluga, scmemoryexpandhpc, donglicxl}.
Compute Express Link (CXL)~\cite{introtocxl} is one of the major proposed memory expansion interconnects, with DRAM-based memory expansion devices being deployed in real-life use cases~\cite{SamsungCMM-D,SKhynixCXL2025}.
With the expansion into CXL, a branch of new CXL memory devices in CXL-SSDs has emerged, proposing memory semantic data access with a small DRAM cache backed with a large pool of NAND in one device~\cite{jung2022cxlssd, cmmhpmem, cmmhhotstorage}.
Ongoing research into CXL-SSD designs aims to realize the potential to deliver near CXL memory speeds at NAND SSD level capacities~\cite{zhang2025skybyte, blocktobyte, cxlssdexpand}.

One of the largest roadblocks for CXL-SSD research is the limited accessibility of real-life hardware to the general community to perform evaluations and build new CXL-SSD designs.
To answer these concerns, many emulators have been proposed, with various limitations including memory trace reliance, limited full-stack software incorporation, and lack of design extensibility~\cite{opencxd, wang2025cxlsim}. Cylon represents the state of the art in CXL-SSD emulation, enabling real-time workload execution with an emulated CXL-SSD device using QEMU, a virtual machine (VM)~\cite{cylon}.
However, the usage of VMs incurs non-trivial amounts of latency and overhead when interception of memory I/O is required to incorporate NAND latency for DRAM misses. 
While this overhead is less noticeable with slower NAND, high performance NAND with lower latencies used in CXL-SSD studies~\cite{blocktobyte, cxlssdexpand} may suffer due to overheads being larger than the designed simulated latency. 

To overcome this limitation, we present \tbdfixed{}, a CXL-SSD emulation platform that runs directly on the host operating system via a kernel module. 
By removing the VM from the equation entirely, \tbdfixed{} reproduces the configured timing parameters more faithfully.

\label{sec:intro}

\section{Problem Definition}
\label{sec:motiv}

In this section, we go over the weaknesses of a CXL-SSD emulation environment like Cylon that is reliant on a VM. 
Namely, we reveal the excessive overhead that VMs bring in median and tail miss latency that harms emulation accuracy.

\begin{figure}[!h]
    \centering
    \vspace{-6pt}
    \includegraphics[width=\linewidth]{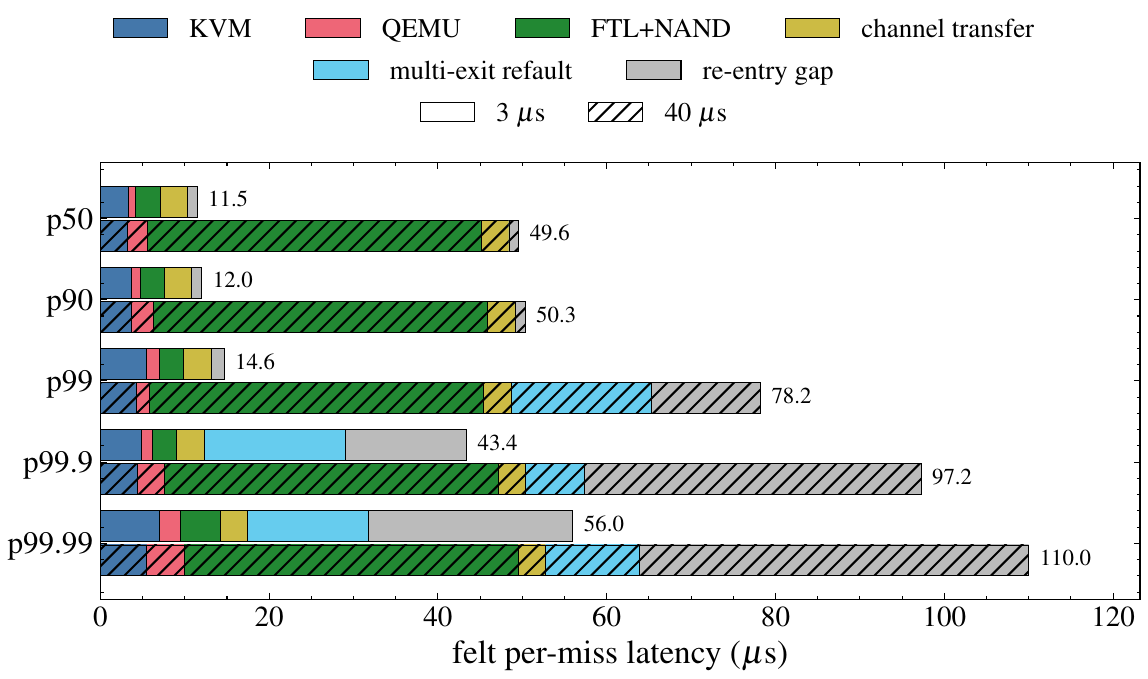}
    \caption{Breakdown of Cylon's DRAM miss latency across two NAND latency profiles.}
    \label{fig:motiv_breakdown}
    \vspace{-6pt}
\end{figure}

\label{sec:motiv_vm_overhead}

To determine the emulation characteristics of the CXL-SSD in Cylon, we run microbenchmarks in the form of a single threaded pointer chase with MIO~\cite{miobench} whose working set is 1.5$\times$ the DRAM cache (1.5$\times$ WSS), using the evaluation platform as specified in Table \ref{tab:system-config}. 
From these benchmarks, we categorize the contributors to the measured miss latency, and break down the results in tail percentiles in Fig.~\ref{fig:motiv_breakdown}. The results reveal two critical issues of Cylon in emulating a CXL-SSD.

\noindent\textbf{KVM/QEMU overhead in emulation.}
A key contributor of latency from using Cylon is the VM mechanisms that underlie the emulation platform. Figure \ref{fig:motiv_breakdown} breaks down the sources of latency into two categories: KVM and QEMU.

\squishlist
\item \textbf{KVM}: Time from host machine's kernel operations that occur from both VM exit and entry due to a DRAM miss. Operations included here are the extended page table (EPT) walk, \texttt{ioctl} syscall operations, MMIO operations, and x86 instruction execution from the EPT trap.

\item \textbf{QEMU}: Actual QEMU operations that pertain to operating the VM. These include the big QEMU lock (BQL) acquisition, memory topology routing, and copy operations from CXL-SSD emulation. 

\squishend

Combined, these two aspects of latency consistently add up to over 3\us{} in all latency setups, appearing in both median and tail latency values at a minimum of 4.15\us{}.
This overhead alone exceeds the read latency of high performance NAND such as Samsung Z-NAND~\cite{z-nand}. 
Therefore, a persistent accuracy issue across Cylon's emulation can be identified, clearly emphasizing the shortcomings of VM-reliant emulation for CXL-SSDs with low NAND latencies.

\noindent\textbf{Additional latency from VM exit overhead.}
Outside of expected (FTL, NAND, channel) and direct VM machinery latency, another major overhead concern is the VM exit overheads.
Such overheads show mostly in tail latency, from P99 to the extreme tails in P99.99.
We identify the following additional contributors to latency from the VM exit.
\squishlist
\item \textbf{Multi-exit refault}: The waiting done to incur NAND latency and the EPT update are independent of each other.
Even if the wait is complete, if the EPT is not marked for direct access, Cylon re-executes the load operation and performs additional VM exits. 

\item \textbf{Re-entry gap}: Unavoidable gaps of time between each VM exit and re-execution from host contention are also a contributing factor in latency.
This is made more evident in the tail latency regions where multi-exits are more common.
An access that misses the direct mapping on its first exit waits out the remainder partly in repeated exits and partly in such gaps.
\squishend

Such high overheads in extreme tail latencies from VM usage are documented in prior studies, and pose concerns for using VM-based emulation in CXL-SSD studies~\cite{nvmevirt}. 
Current CXL-SSD designs with a small DRAM cache in comparison to the large NAND capacity make the accurate portrayal of DRAM misses critical~\cite{cmmhpmem, cmmhhotstorage}.
A small DRAM cache makes misses more likely to occur, thus elevating the impact of extreme tail latencies.
Cylon's failure to portray the configured latencies reveals a gap in current emulation fidelity, thus motivating a new approach for CXL-SSD emulation.

\section{Design of \tbdfixed{}}
\label{sec:design}

\subsection{Component Overview}
\label{subsec:components}
\begin{figure}[!h]
    \centering	
    \vspace{-10pt}
    \includegraphics[width=\linewidth]{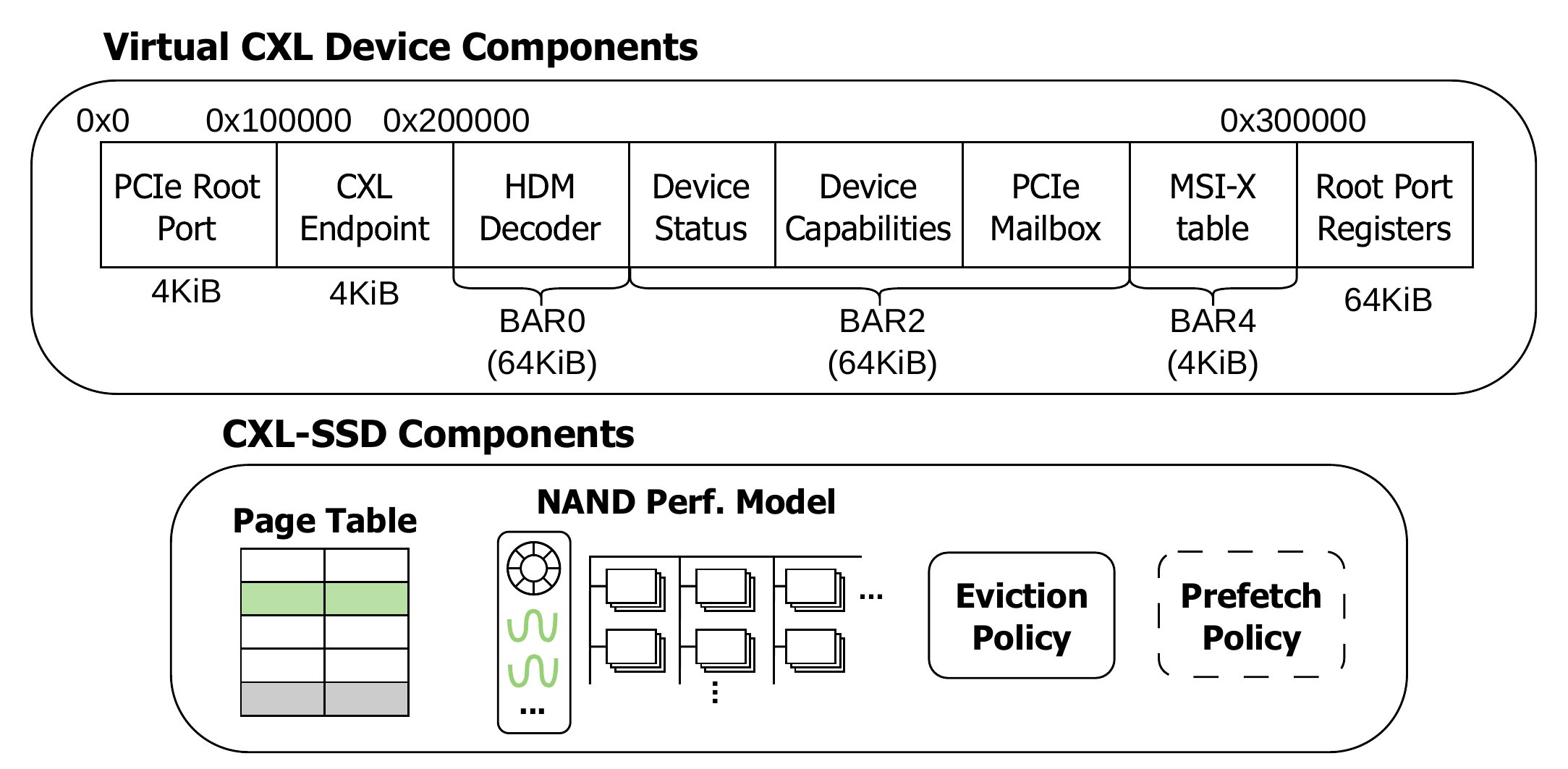}
    \vspace{-3pt}
    \caption{Components of \tbdfixed{} to ensure functionality.}
    \vspace{-6pt}
    \label{fig:design_components}
\end{figure}

Fig.~\ref{fig:design_components} outlines the components that drive \tbdfixed{}'s functionality.
They are divided into components for initializing the actual CXL-SSD device via the kernel module, and components for the operation and features of the CXL-SSD.

\noindent\textbf{Components for CXL device registration.}
In order to identify the device created by \tbdfixed{}'s kernel module as a CXL memory device, the relevant device metadata for the PCIe device creation must be created. 
This metadata is stored in a small reserved space of DRAM used for device emulation on the host, and includes the CXL endpoint information, PCIe BAR regions, the MSI-X table, and CXL component and device registers.

\noindent\textbf{Components for CXL-SSD operation management.}
To drive the NAND read, write, and erase latencies of I/O between the DRAM and NAND region, \tbdfixed{} employs a NAND performance model.
Additional components include the eviction and prefetch policies that a CXL-SSD device design may want to implement.
\tbdfixed{} offers a variety of eviction policies that determine what pages to evict from DRAM to NAND, such as FIFO, LIFO, and CLOCK.
For prefetch policies, \tbdfixed{} offers a general next $N$ page background prefetch policy for exploiting spatial locality.

\tbdfixed{} is flexible and modifiable with customizable eviction and prefetch policies to suit a specific workload's needs.
To support these features, \tbdfixed{} keeps per-page metadata for the CXL-SSD as seen in Table~\ref{tab:metadata}.
Each page's entry holds its state, dirty bit, CLOCK bit, and pin count, its position in the FIFO/LIFO order, and a wait queue for accesses waiting on its state transitions. 
Bitmaps are used to track each page's residency, validity, allocation, and prefetch status.
The data stored here is not part of the DRAM region exposed as usable memory space, and is considered as a separate reserved memory region.
Future CXL-SSD designs with \tbdfixed{} can share this memory region with the host system to facilitate communication with the device for hardware-software co-design studies.
\begin{table}[h]
  \centering
  \small
  \caption{\tbdfixed{} per-page metadata for the 96\,GiB device
           (23.8\,M tracked 4\,KiB pages).}
  \label{tab:metadata}
  \setlength{\tabcolsep}{4pt}
  \resizebox{\linewidth}{!}{
  \renewcommand{\tabularxcolumn}[1]{m{#1}}   %
  \begin{tabularx}{\columnwidth}{@{}>{\raggedright\arraybackslash}Xrr@{}}
    \toprule
    Structure & Per page & Total \\
    \midrule
    Page entry (state, CLOCK bit, pin count, dirty bit, FIFO/LIFO order, wait queue) & 48\,B   & 1,091.4\,MiB \\
    Bitmaps (resident, valid, alloc, prefetch)  & 5\,bits & 14.2\,MiB \\
    \midrule
    Total            & 48.6\,B & 1,105.6\,MiB \\
    \bottomrule
  \end{tabularx}
  }
\end{table}

\subsection{Operation Flow of \tbdfixed{}}
\begin{figure}[!h]
    \centering	
    \vspace{-10pt}
    \includegraphics[width=0.75\linewidth]{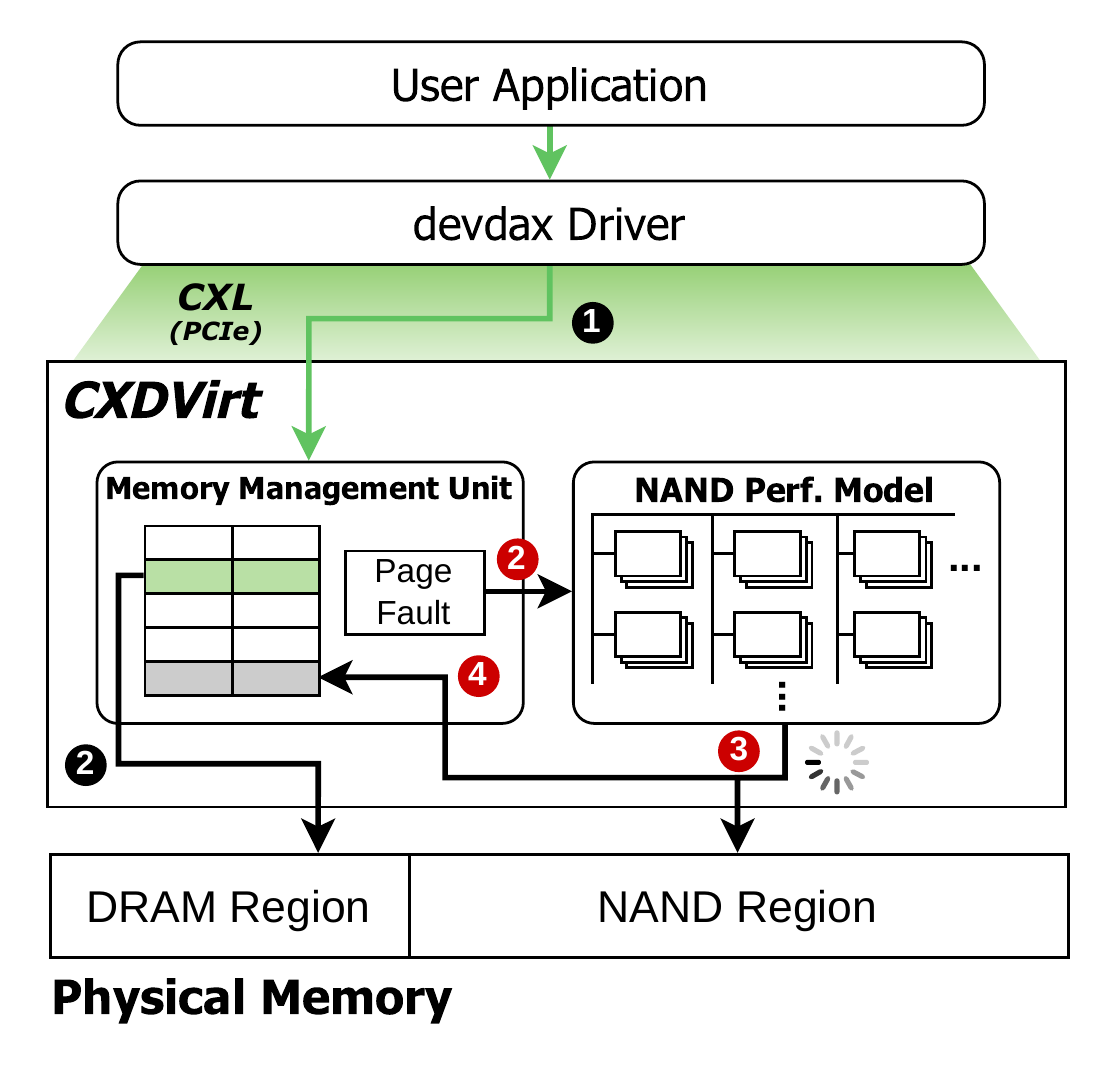}
    \vspace{-7pt}
    \caption{Overview of \tbdfixed{}'s execution flow.}
    \vspace{-16pt}
    \label{fig:design_overview}
\end{figure}
\tbdfixed{}'s key method of enabling the CXL-SSD emulation is the kernel's page table and page-fault handler machinery. \tbdfixed{} uses the entire reserved physical memory region as capacity for the purposes of emulation.

Fig.~\ref{fig:design_overview} illustrates this with the overall architecture and execution flow. 
\tbdfixed{} starts with the user application accessing the emulated CXL-SSD device via the devdax driver (\CircledTextBlack{1}).
Here, the execution flow diverges on whether the data is in DRAM (DRAM hit) or in NAND (DRAM miss), determined by using the page table.
If a DRAM hit occurs, the memory management unit (MMU) directly translates and accesses the location of the data in DRAM (\CircledTextBlack{2}). This enables data access at pure memory access speed without any additional interception.

On the other hand, a DRAM miss triggers a custom page fault implemented using the devdax device (\CircledTextRed{2}).
Instead of looking for data in storage like a normal page fault, the page fault here is used as an interception point for memory accesses.
By intercepting at this point, \tbdfixed{} uses a NAND performance model to determine the incurred NAND I/O time in the CXL-SSD. 
After waiting for the calculated NAND latency, access to the data in physical memory finally occurs (\CircledTextRed{3}).
At the same time, the page table is updated to mark said page is now present in memory, allowing for native memory access times (\CircledTextRed{4}) for future access to the same page.

\noindent\textbf{Flush/retrieve between regions.}
To avoid the overhead of data movement and \texttt{memcpy}, \tbdfixed{} sets no static DRAM or NAND regions, nor does it move data between the two regions in both flush and retrieve operations.
Rather, the distinction of the data's location is determined by its page table entry (PTE).
Flush from DRAM to NAND removes the PTE to incur a page fault on its next access.
NAND writes are only issued if the page is dirty, and can be issued either by the foreground faulting thread or \tbdfixed{}'s background I/O workers depending on whether the flush is to be performed synchronously or asynchronously.
Data retrieval from NAND to DRAM installs the PTE to allow direct data access via the MMU. 
NAND reads are issued in the foreground, or in the background if the retrieval was from prefetch.

\begin{figure}[!h]
    \centering	
    \vspace{-10pt}
    \includegraphics[width=\linewidth]{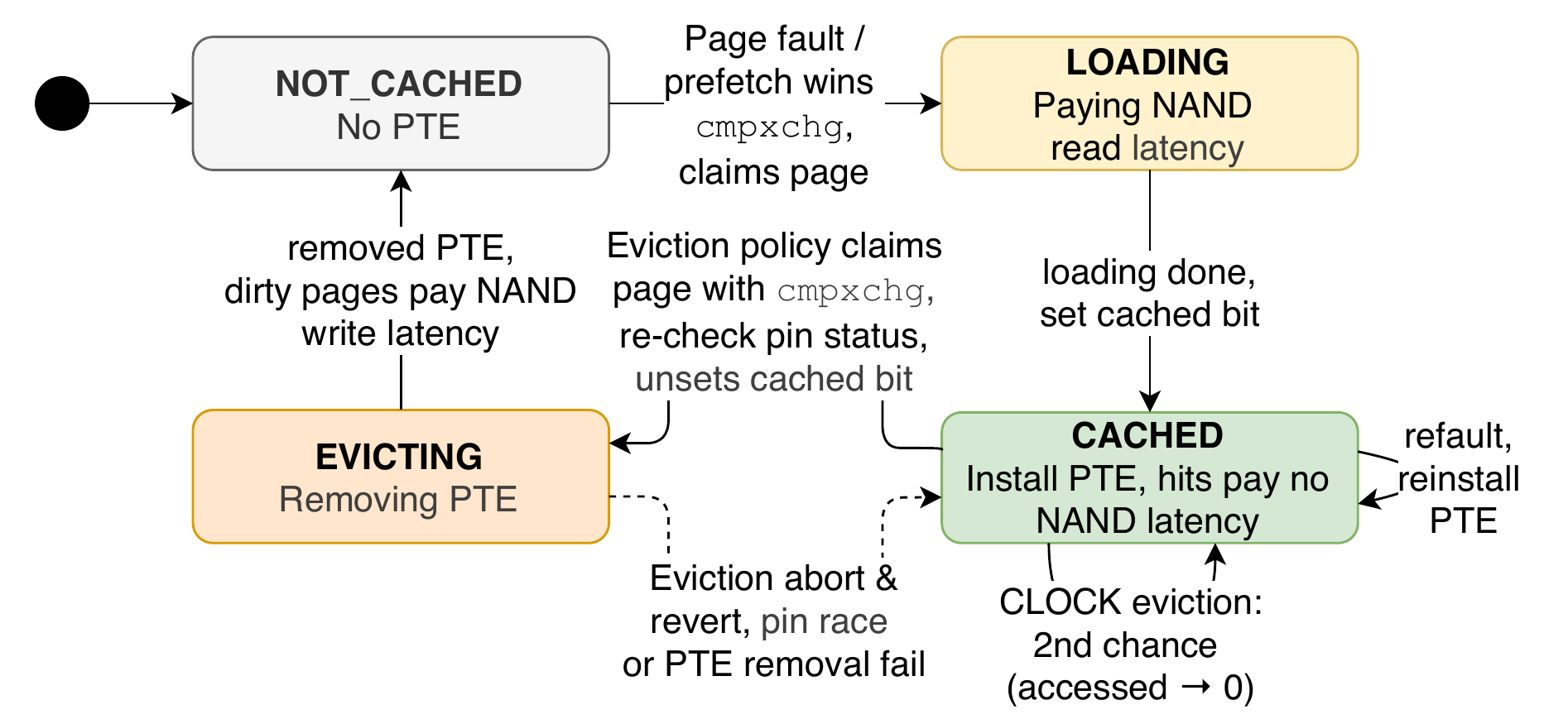}
    \vspace{-6pt}
    \caption{State machine diagram in \tbdfixed{}.}
    \vspace{-8pt}
    \label{fig:design_state}
\end{figure}

\subsection{Page state management in \tbdfixed{}}

A state machine is used to represent the status of pages currently stored in \tbdfixed{} as seen in Fig.~\ref{fig:design_state}. 
Each page has information on its state saved and accessed in its page metadata entries  (\S\ref{subsec:components}, Table~\ref{tab:metadata}).
Each page is in exactly one of four states as seen below.
\squishlist
\item \textbf{NOT\_CACHED}: 
Page is located in the NAND region, and requires a NAND read to the DRAM region for usage.

\item \textbf{CACHED}: Page is located in the DRAM region.

\item \textbf{EVICTING}: Page is in the process of being written from DRAM to NAND if said page is dirty.

\item \textbf{LOADING}: Page is in the process of being read from NAND to DRAM.
\squishend
Each state change is claimed with a per-page atomic compare-and-swap, with no device-wide lock held while a miss waits out its NAND latency. 
Therefore, misses to different pages can proceed in parallel, whereas Cylon serves every miss under QEMU's global lock.
Concurrent accesses to a page in LOADING or EVICTING sleep on its wait queue, so each fill is charged once, and a pin count keeps eviction from removing a page whose PTE is being installed.

\section{Evaluation}
\label{sec:eval}

\subsection{Implementation \& evaluation setup}
\begin{table}[!h]
\vspace{-6pt}
\centering
\footnotesize
\setlength{\aboverulesep}{0pt}
\setlength{\belowrulesep}{0pt}
\setlength{\extrarowheight}{1pt}
\renewcommand{\arraystretch}{0.95}
\caption{Specifications of evaluation platform.}
\label{tab:system-config}
\begin{tabular}{@{}lcc@{}}
\toprule
            & Cylon & \tbdfixed{} \\
\midrule
CPU         & \multicolumn{2}{c}{Dual Xeon Gold 5218R, 40\,C/80\,T} \\
NAND        & \multicolumn{2}{c}{Z-NAND~\cite{z-nand} SLC, 8\,ch $\times$ 16\,way} \\
NAND timing & \multicolumn{2}{c}{$t_\text{R}$/$t_\text{PROG}$/$t_\text{BERS}$ = 3/100/1000\,\us{}} \\
\midrule
OS          & Ubuntu 22.04.5 & Ubuntu 20.04.6 \\
Kernel      & Linux 6.4.6    & Linux 6.18.5 \\
Workloads   & VM, 8 vCPUs    & host \\
Memory      & 96\,GiB        & 282\,GiB \\
\bottomrule
\end{tabular}
\end{table}
We implement \tbdfixed{}\footnote{\url{https://github.com/hschung1652/cxdvirt}} through a modified version of NVMeVirt~\cite{nvmevirt}, with the evaluation system as specified in Table \ref{tab:system-config}.
PCI device creation and NAND performance modeling are taken from NVMeVirt, with modifications made for CXL device realization and CXL-SSD device semantics.
A custom kernel is implemented to expose the required functions to the \tbdfixed{} module for the CXL device creation and page table manipulation.
A 96GiB region on a remote NUMA node serves as the CXL-SSD's physical storage medium, of which 4.8GiB acts as the DRAM cache unless otherwise stated; said remote placement enables incurring non-local memory access latency to the backing medium.
All workload sizes are given as working set size (WSS) ratio to the DRAM cache capacity.

\subsection{Microbenchmarks: Device latency characteristics}
\label{sec:micro}
\begin{figure}[h]
    \centering
    \vspace{-10pt}
    \includegraphics[width=\linewidth]{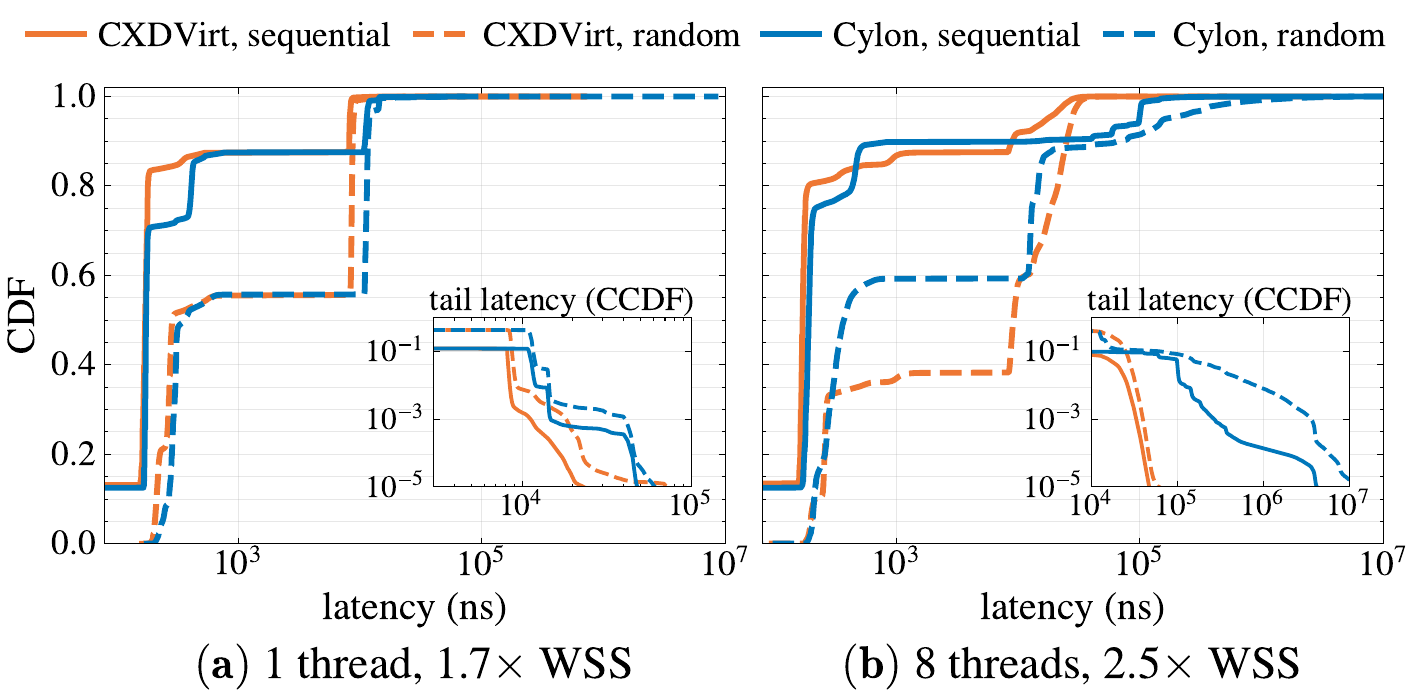}
        \caption{Access latency CDFs for sequential and random pointer chasing across Cylon and \tbdfixed{}.}
    \label{fig:eval_cdf}
    \vspace{-10pt}
\end{figure}

\noindent\textbf{Overall performance characteristics:} 
Fig.~\ref{fig:eval_cdf} shows the distribution of Cylon and \tbdfixed{} latencies running MIO in a variety of settings.
Cylon and \tbdfixed{} both show the bimodal hit/miss profile expected of a CXL-SSD, but \tbdfixed{}'s misses sit closer to the configured device latency ($t_\text{R}$ plus the 4\,KiB channel transfer) of 6.2\us{}.
At one thread, \tbdfixed{}'s curves show only the two steps with the miss step 1\us{} wide at 8--9\us{}, while Cylon's sequential curves add a third step for the access after each miss and place their misses at 10--12\us{}. 
Under eight threads, contention widens \tbdfixed{}'s misses to 8--34\us{}, whereas Cylon's spread from 12\us{} to as much as 1.7\,ms.
On average, a miss at eight threads takes 1.9--2.0$\times$ as long as at one thread on \tbdfixed{}, but 8.0--9.2$\times$ on Cylon.
The complementary CDF (CCDF) inset graphs take a closer look at tail latencies, which reflect DRAM miss distributions. 
Here, \tbdfixed{} is lower at every level from $10^{-1}$ down to $10^{-4}$ at one thread, and above P90 at eight threads.
At eight threads, Cylon's accesses take 4.6--25$\times$ longer at P99 and 8--91$\times$ longer at P99.9.

Another point of interest is the lead in Cylon's random curve in Fig.~\ref{fig:eval_cdf}(b) until P88. 
This is not due to lower emulation latency, as Cylon's misses take 6$\times$ longer on average, but to its miss path through QEMU.
With uniformly random access, at most $\frac{1}{\text{WSS}}$ of accesses can hit, which is 40\% in (b). 
\tbdfixed{}'s plateau sits just below this at 0.38, while Cylon's rises well above it to 0.59.
Exceeding the bound requires uneven thread progress, which is caused by QEMU's global lock. 
Cylon serves each miss, NAND wait included, while holding this lock, and 71\% of the time the lock then passes to the same vCPU's next miss.
The lock holder's misses are thus served back to back in about 13\us{}, forming the steep rise from
0.59 to 0.88, while the other vCPUs' misses wait and form the slope above P88.
Running ahead also keeps the holder's region cached, which raises Cylon's hit plateau: its threads hit 52--74\% and spend 62--177\,s in the run, while every \tbdfixed{} thread hits 38\% and spends 33\,s.

\begin{figure}[h]
    \centering
    \includegraphics[width=\linewidth]{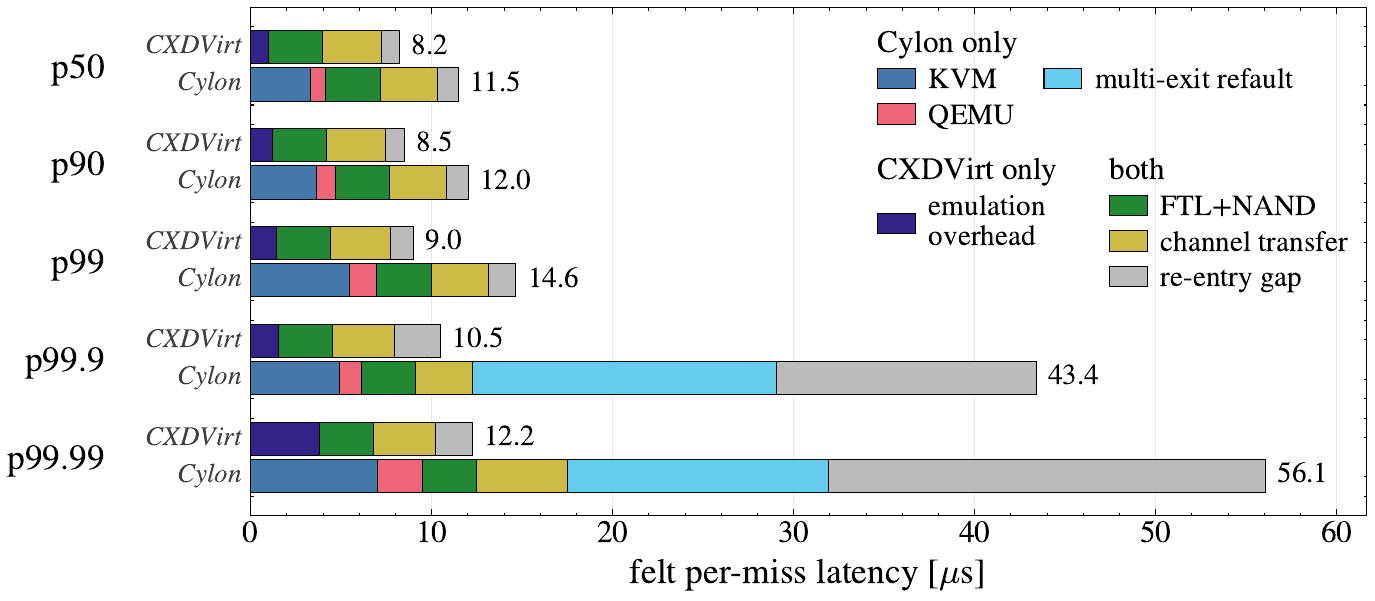}
        \caption{DRAM miss latencies of \tbdfixed{} and Cylon across a variety of percentiles.}
    \label{fig:eval_drammiss}
    \vspace{-10pt}
\end{figure}

\noindent\textbf{DRAM miss breakdown:} 
Fig.~\ref{fig:eval_drammiss} shows a closer study of DRAM miss latencies from a single-threaded sequential pointer chase workload at 1.5$\times$ WSS.
The results reveal why \tbdfixed{} is able to achieve lower DRAM miss latencies over Cylon as seen in Fig.~\ref{fig:eval_cdf}.
Up to P99 of DRAM misses, \tbdfixed{} maintains a 1.4$\times$ to 1.6$\times$ reduction in latency.
Entering the extreme tail widens this gap further, with 4.1$\times$ and 4.6$\times$ for P99.9 and P99.99 respectively.
This reduction is due to \tbdfixed{}'s ability to eliminate the VM-specific overheads identified in \S\ref{sec:motiv}.
Against the configured miss latency of 6.2\us{}, the application observes a 2.0--2.7\us{} overhead on \tbdfixed{} up to P99 and 4.1\us{} at P99.9, against 5.3--8.5\us{} and 37.2\us{} on Cylon.
\tbdfixed{} therefore only leaves the cost of the page fault itself between the configured and the observed latency, making the device model the the main determinant of what an application observes.

\subsection{Macrobenchmarks: real-life workloads with \tbdfixed{}}
For real-life workloads, we run the write-intensive Splash-4 OCEAN workload with eight threads, and the read-intensive YCSB-C workload with Redis in a single thread across three WSS ratios~\cite{splash4, ycsb, redis}. 
For the OCEAN workload, the DRAM region's size is adjusted to meet the specified WSS ratio to the workload's size of 13.9GiB.
A remote NUMA DRAM baseline was also performed to compare with the 0.35 WSS result, which represents a CXL-SSD workload that fits entirely within the DRAM region with no NAND operations involved.

\begin{figure}[h]
    \centering
    \includegraphics[width=\linewidth]{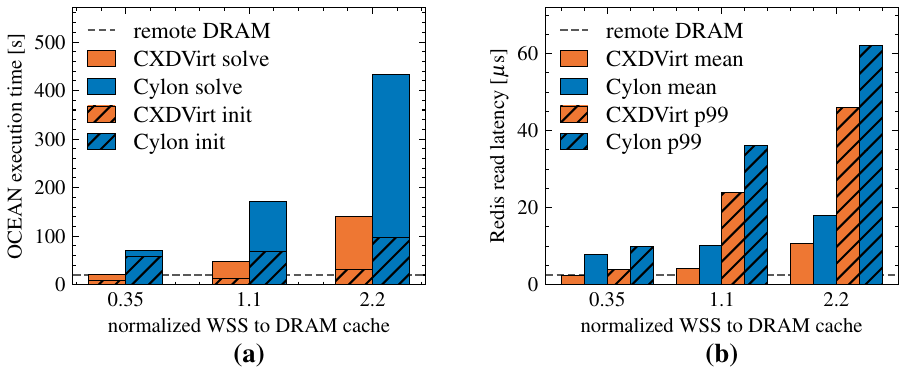}
        \caption{Results of \tbdfixed{} and Cylon for (a) Splash-4 OCEAN and (b) Redis YCSB-C.}
    \label{fig:eval_macro}
    \vspace{-6pt}
\end{figure}
Fig.~\ref{fig:eval_macro}(a) shows the Splash-4 results in total execution time, where \tbdfixed{} maintains a 3.06$\times$ to 3.51$\times$ execution time reduction over Cylon. Cylon's overhead is clearly visible at 0.35 WSS, where no NAND operations occur due to the small WSS. While \tbdfixed{}  stays within 5\% of the remote DRAM backing the emulation, Cylon faces a 3.6$\times$ increase, indicating emulation machinery overhead. This overhead continues up to 2.2 WSS, where \tbdfixed{} sees 1.1--2.6\us{} per miss overhead excluding remote-DRAM time and modeled NAND latency against 15.8--17.3\us{} on Cylon.

Fig.~\ref{fig:eval_macro}(b)'s Redis results show a similar pattern to Splash-4.
At 0.35 WSS, \tbdfixed{} meets remote DRAM latency, while Cylon is 3.1$\times$ slower, its mean at least 5\us{} above \tbdfixed{}'s at all three ratios.
This extra overhead is especially noticeable in P99 latencies from DRAM miss latency, with a 1.5$\times$ and 1.35$\times$ increase at 1.1 and 2.2 WSS ratios respectively. 
The overhead also makes Cylon 1.9$\times$ less sensitive to WSS growth: from 0.35 to 2.2 WSS, \tbdfixed{} reports mean latency rising 4.38$\times$ while Cylon only 2.31$\times$.

\begin{figure}[h!]
    \centering
    \includegraphics[width=\linewidth]{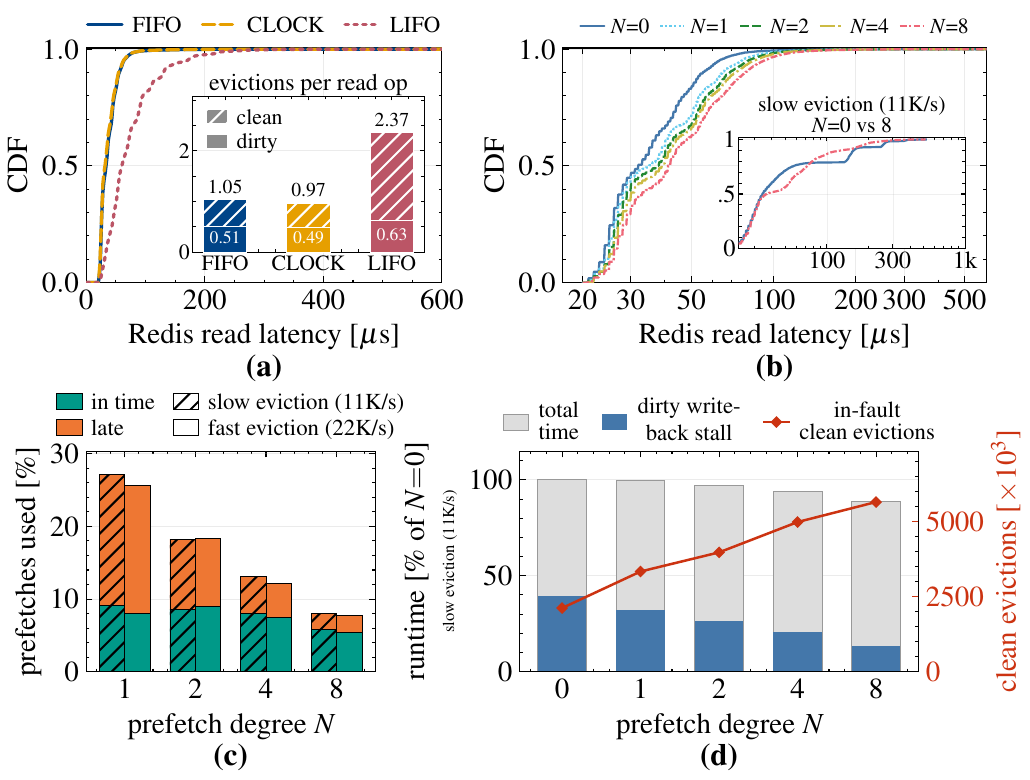}
        \caption{Eviction (a) \& prefetch (b)--(d) policy study results.}
    \label{fig:eval_casestudy}
    \vspace{-8pt}
\end{figure}

\noindent\textbf{Eviction \& prefetch policy study:}
As a case study, we conduct eviction and prefetch policy studies and show the difference in performance characteristics in Fig.~\ref{fig:eval_casestudy}. 
Both studies perform the Redis YCSB-C benchmark with a single thread and maintain a WSS ratio of 2.65 to stress the DRAM cache.

Fig.~\ref{fig:eval_casestudy}(a) covers the eviction policy of the study.
The CDF in Fig.~\ref{fig:eval_casestudy}(a) shows that while FIFO and CLOCK show near identical latency characteristics, LIFO shows significantly worse results, with 1.8--1.9$\times$ higher read latency medians over FIFO and CLOCK.
The bar graph shows dirty evictions stay at 0.49--0.63 per read for all three policies, indicating that workload, more than the eviction policy, determines the dirty rate of pages.
LIFO's 3.2--3.6$\times$ higher clean eviction rate also shows that its cost in read latency is in repeated refetches over extra writeback traffic.

Fig.~\ref{fig:eval_casestudy}(b)--(d) covers the prefetch study, showing next-$N$ prefetch as ineffective for the Redis workload: read latency rises monotonically with $N$ at every percentile in Fig.~\ref{fig:eval_casestudy}(b).
The extra retrievals amplify evictions by 6.1$\times$ with the Redis thread paying both the prefetch and the eviction machinery.
Prefetch helps only once background eviction is slowed, and only in the upper tail with a 42\% reduction at $N{=}8$'s P80.
This gain is not better hit rate, as Fig.~\ref{fig:eval_casestudy}(c) shows fewer than 10\% of prefetches arrive in time, and the two eviction rates agree to within 1.5 points on both prefetch accuracy and timeliness.

Fig.~\ref{fig:eval_casestudy}(d) resolves the question of how slower eviction takes advantage of prefetch.
With background eviction lagging behind, evictions in demand faults are more common, where it is susceptible to a dirty page writeback.
Meanwhile, prefetched pages are often clean, so in-fault evictions can pick clean prefetched pages as victims instead of dirty ones, reducing the amount of writeback stall. 
Hence, the latency reduction comes from preventing writebacks, not from prefetched data being used.
The case study shows the level of detail \tbdfixed{} offers for CXL-SSD machinery studies, and resolves subtle policy questions for a device with the configured timing parameters.

\section{Discussion \& Future Work}

Following Cylon and prior CXL emulation studies \cite{pond, hu2026aquiferhierarchicalmemorypooling, cylon}, \tbdfixed{} represents CXL memory timings with remote NUMA access latencies---a platform limitation and a source of potential evaluation inaccuracy.
Where available, real CXL memory can instead serve as the backing medium, capturing true physical CXL characteristics while incurring emulated NAND I/O latencies on DRAM misses.
\tbdfixed{}'s lower emulation overhead makes it a compelling basis for such a solution, which we leave to future work.

\section{Conclusion}

We present \tbdfixed{}, a new CXL-SSD evaluation platform that circumvents the overhead of VMs.
Through \tbdfixed{}, we hope to make CXL-SSD research more accessible by enabling the wider community to contribute new optimized designs of both general and specialized use cases for the emerging memory device.

\section*{Acknowledgments}
{
This work was supported by the National Research Foundation of Korea (NRF) grant funded by the Korea government (MSIT) (RS-2025-00564249 and RS-2024-00453929) and by the Institute for Information \& Communications Technology Promotion (IITP) grant funded by the Korea government (MSIT) (No. RS-2025-02215256).
}

\scriptsize 
{%
  \bibliographystyle{ieeetr}
  \bibliography{paper}
}

\end{document}